\documentstyle[psfig,proceedings]{crckapb} 

\def\Msun{\ifmmode{~M_\odot}\else$M_\odot$~\fi}
\def\rsun{\ifmmode{~r_\odot}\else$r_\odot$~\fi}
\def\kms{\ifmmode{$~km\thinspace s$^{-1}~}\else km\thinspace s$^{-1}~$\fi}
\def\ga{\mathrel{\mathchoice {\vcenter{\offinterlineskip\halign{\hfil
$\displaystyle##$\hfil\cr>\cr\noalign{\vskip1.5pt}\sim\cr}}}
{\vcenter{\offinterlineskip\halign{\hfil$\textstyle##$\hfil\cr>\cr
\noalign{\vskip1.0pt}\sim\cr}}}
{\vcenter{\offinterlineskip\halign{\hfil$\scriptstyle##$\hfil\cr>\cr
\noalign{\vskip0.5pt}\sim\cr}}}
{\vcenter{\offinterlineskip\halign{\hfil$\scriptscriptstyle##$\hfil
\cr>\cr\noalign{\vskip0.5pt}\sim\cr}}}}}
\def\la{\mathrel{\mathchoice {\vcenter{\offinterlineskip\halign{\hfil
$\displaystyle##$\hfil\cr<\cr\noalign{\vskip1.5pt}\sim\cr}}}
{\vcenter{\offinterlineskip\halign{\hfil$\textstyle##$\hfil\cr<\cr
\noalign{\vskip1.0pt}\sim\cr}}}
{\vcenter{\offinterlineskip\halign{\hfil$\scriptstyle##$\hfil\cr<\cr
\noalign{\vskip0.5pt}\sim\cr}}}
{\vcenter{\offinterlineskip\halign{\hfil$\scriptscriptstyle##$\hfil
\cr<\cr\noalign{\vskip0.5pt}\sim\cr}}}}}
\begin{opening}
\title{KINEMATICS AND DYNAMICS OF \protect\\
       THE GALACTIC STELLAR HALO}

\author{JESPER SOMMER-LARSEN}
\institute{Theoretical Astrophysics Center\\
           Juliane Maries Vej 30\\
	DK-2100    Copenhagen {\O},  Denmark}

\end{opening}

\runningtitle{KINEMATICS AND DYNAMICS OF THE STELLAR HALO}

\begin{document}


\begin{abstract}
The structure, kinematics and dynamics of the Galactic stellar halo are reviewed including
evidence of substructure in the spatial distribution and kinematics of halo stars.  
Implications for galaxy formation theory are subsequently discussed; in particular
it is argued that the observed kinematics of stars in the outer Galactic halo can be used as
an important constraint on viable galaxy formation scenarios.
\end{abstract}

\section{Introduction}
Although the stellar halo accounts for at most a few percent of the luminous mass of
the Galaxy, it plays a crucial role in studies of the Galaxy's
formation, evolution, and present-day structure.  The halo has long
been considered the Galaxy's oldest component. Age estimates can be obtained
for its most conspicuous constituent, the metal-weak globular
clusters, as well as for individual metal-weak stars (with less certainty).  
Thus the dynamical and chemical state of the luminous halo
population provides information on the formation of large disk galaxies such
as the Milky Way.

\section{Is there substructure in the halo ?}
In the currently favoured hierarchical galaxy formation scenarios big disk galaxies
like the Milky Way are built up through merging of smaller subsystems. In particular
it is thought that the Galactic halo was formed and, to some extent, still is being formed
by accretion of many small satellites.

Likely examples of ongoing {\it stellar} accretion are the Sagittarius dwarf 
(Ibata, Gilmore \& Irwin 1994) and the Magellanic Stream (Majewski {\it et al.} 1998). 
Furthermore one would generally expect strong
substructure in the distribution of halo stars in phase-space due to past accretion
events (e.g., Johnston, Spergel \& Hernquist 1995). 
This substructure is difficult to detect due to the
very low number density of halo stars (``observational shot-noise'') and the physical
effect of phase-mixing, but accreted systems with extreme kinematics should be
detectable --- one very likely example of this is the moving group of Majewski, Munn \&
Hawley (1994). The blue metal-poor stars (BMPs) of Preston, Beers \& Shectman (1994) 
may well be
another example of a population of halo stars accreted in one (major) event --- see
also Carney {\it et al.} (1996).

\section{The local halo velocity ellipsoid and the structure of the inner halo}
The inner halo, defined here as the part of the halo inside of or about at the solar distance
from the center of the Galaxy ($r \la r_{\odot} \simeq$ 8 kpc) consists of at least 
two components: 
the flat and the round halo. The ratio of the local density of flat to round halo is not
well determined ranging from 0.5--1 (Sommer-Larsen \& Zhen 1990) to $\sim$ 4--8
(Kinman, Suntzeff \& Kraft 1994, Hartwick 1987). The velocity ellipsoid of local halo stars 
has been determined by Beers \& Sommer-Larsen (1995). They used radial velocities
of stars selected without kinematic bias, making the result very robust, also to distance
errors. For almost 900 stars with [Fe/H] $< -1.5$ they find a velocity ellipsoid for
local halo stars of 
$(\sigma_r, \sigma_{\phi}, \sigma_{\theta}) = (153 \pm 10,
93 \pm 18, 107 \pm 7)$ \kms in spherical polars. 
This velocity ellipsoid 
is fairly radially elongated, but is still characterized by a quite large vertical (and horizontal) 
tangential velocity dispersion of $\sim$ 100 \kms. It follows from the tensor
virial theorem that the flat component can not be both very flat {\it and}
locally dominant --- see Sommer-Larsen \& Christensen (1989).

Carney {\it et al.} (1996) analyzed two classes of halo stars from their local sample:
The ``low'' halo stars have $\langle|z_{max}|\rangle \le$ 2 kpc and the ``high'' halo stars
have $\langle|z_{max}|\rangle \ge$ 5 kpc, where $\langle|z_{max}|\rangle$ is 
the typical distance
a star reaches from the plane of the disk --- as the Carney {\it et al.} stars are all local
$z_{max}$ was calculated from the observed space motions by orbit integration. 
These two classes of halo stars can be
seen as representative of the flat and round halo respectively. The ``low'' halo
is characterized by a slight prograde average rotation \
$\langle v_{\phi}\rangle \simeq 20 \pm 15$
\kms with respect to the Galactocentric restframe assuming a circular speed at the
solar distance from the Galactic center of $v_{\odot}$ = 220 \kms. The ``high'' halo, on
the other hand, is characterized by a net retrograde rotation $\langle v_{\phi}\rangle \simeq 
-50 \pm 15$ \kms. The latter result was first found by Majewski (1992) from proper
motions of halo stars situated at the north galactic pole (NGP) and at least 5 kpc
from the plane of the disk. The results above are further indications of kinematic
substructure in the halo. Surprisingly a similar retrograde net rotation has not been
detected so far for halo stars {\it in situ} at $|z| >$ 5 kpc at the SGP (Beers, private
communication). As the retrograde halo
stars are found both at high $z$ at the NGP and locally these stars should be
well mixed, so from stellar dynamics considerations it follows that such stars
{\it should} be found at the SGP in the future. 

Martin \& Morrison (1998) find
in an interesting recent study of the kinematics of local ($d \la$ 1 kpc)
RR-Lyrae stars with accurate 3-D velocities that $\sigma_r = 193 \pm 15$ \kms
for the halo stars in their sample. As noted by the authors this differs from the 
result of Beers \& Sommer-Larsen (1995) for halo stars in general by more than 2 $\sigma$ (see above).
A possible explanation of this difference is that the spatial distribution of the
RR-Lyrae halo stars differs from that of the halo stars in general:
Approximating the spatial distributions by power-laws $\rho \propto r^{-\alpha}$,
then $\alpha_{RR} = 3.2 \pm 0.1$, whereas for the halo stars in general $\alpha_{HALO}
\simeq 3.5$ --- see, e.g., Preston, Shectman \& Beers (1991). 
Pushing $\alpha_{RR}$ to 3.1 and assuming the Beers \& Sommer-Larsen value for
$\sigma_r$ of halo stars in general
it follows from the Jeans equation that one would expect 
$\sigma_r \simeq 179 \pm 12$ \kms for the RR-Lyrae halo stars, in good
agreement with the results of Martin \& Morrison.

Martin \& Morrison used the ``old'' distance scale with $M_V(RR)$ = 0.73 at
[Fe/H] = -1.9. A ``new'' distance scale, based on {\it Hipparcos} trigonometrical
parallaxes of Cepheid variables and subdwarfs, has recently been advocated by
Feast \& Catchpole (1997) and Chaboyer {\it et al.} (1998) resulting in $M_V(RR) \simeq$ 
0.32 at [Fe/H] = -1.9. With this distance scale Martin \& Morrison obtain $\sigma_r \simeq 225 \pm
18$ \kms for their RR-Lyrae halo stars. This differs from the value predicted above
for $\alpha_{RR}$ = 3.1 by more than 2 $\sigma$. Hence, as also noted by 
Martin \& Morrison,
one should perhaps be somewhat cautious about using the ``new'' distance scale, despite its
many merits, like resulting in cosmologically ``reasonable'' globular cluster ages of about 
12 Gyrs, a fairly low value of the Hubble constant etc.

\section{The outer halo}
The outer stellar halo ($r \ga r_{\odot}$) is approximately spherical --- see 
references in Sommer-Larsen {\it et al.} (1997), but also Sluis \& Arnold (1998). 
Several types of stars have been used as tracers of the outer halo: K-giants (e.g,
Ratnatunga \& Freeman 1989), RR-Lyrae stars (e.g., Hawkins 1984) and
blue horizontal branch stars (e.g., Sommer-Larsen {\it et al.} 1997 and references
therein). 
The blue horizontal branch field (BHBF) stars have proven to be particularly useful
tracers of the the outer halo for two main reasons: (a) They are easy to identify in the
halo because of their blue colours and  (b) Using a medium-sized telescope,
spectra of sufficient quality for accurate 
line-of-sight velocity and Balmer line-width determination
can be obtained fairly easily, even for quite distant stars ($d \sim 30-60$ kpc),
because of their intrinsic brightness. Furthermore they seem representative of the
stellar halo in general since the density fall-off of the BHBF stars
in the outer halo can be well approximated by the power-law relation
$\rho (r) \propto r^{-\alpha}, ~\alpha=3.4 \pm 0.3$ --- see Sommer-Larsen,
Flynn \& Christensen (1994).

Sommer-Larsen {\it et al.} (1997) analyzed a sample of almost 700 BHBF stars with good
line-of-sight velocity determinations and situated at Galactocentric distances 
$r \sim$ 7--70 kpc. At distances $d \ga$ 20--30 kpc the line-of sight velocity is close to
being the radial component of the velocity in Galactocentric coordinates. Hence it is
possible to determine the radial velocity dispersion at large Galactocentric distances
from line-of-sight velocities only. Whereas the radial velocity dispersion of
local halo stars is 140--150 \kms $\sigma_r$ is found to drop to about 100 \kms
at large $r$. This kinematic behaviour is modelled in the following simple way
by Sommer-Larsen {\it et al.}:

It is assumed that both the {\it outer} Galactic halo and gravitational potential are
spherically symmetric (moderate departures from this does not affect the outcome
of the analysis in any significant way). The radial velocity dispersion as a function of
$r$ is modelled as 
$$
\sigma_r = \left(\sigma_0^2+\frac{\sigma_+^2}{\pi}\left(\frac{\pi}{2}-{\rm
Arctan}(\frac{r-r_0}{l})\right)\right)^{1/2}.\eqno(1)$$ Adopting this form gives good
flexibility in modelling the decrease in $\sigma_r(r)$ with increasing $r$.
It follows from eq. [1] that $\sigma_0$ is the asymptotic value of the
radial velocity dispersion for $r >> (r_0 + l)$ and that
$\sqrt{\sigma_+^2 + \sigma_0^2}$ approximately is the radial velocity dispersion
in the inner halo ($r \la r_{\odot}$). The physical meaning of the two
scale parameters $r_0$ and $l$ is given in Sommer-Larsen, Flynn \& Christensen (1994) 
and is fairly straightforward.

The rotation curve of the Galaxy is approximately flat to at least $R \simeq$ 20 kpc
(Fich \& Tremaine 1991) and most likely to much larger distances as shown by
Kochanek (1996). Consequently the gravitational potential of the outer halo is
approximated by $$\Phi (r) = V_{c}^{2} \ln(r) ~~, \eqno(2)$$
corresponding to a flat rotation curve  with $v_c(r) \equiv V_c$ = 220 \kms.
Substituting this and eq. [1] into the Jeans equation
$$  \frac{1}{\rho} \frac{d(\rho\sigma_r^2)}{dr} +
\frac{2(\sigma_r^2-\sigma_t^2)}{r} = - \frac{d\Phi}{dr} ~~,  \eqno(3) $$
where $\sigma_t$ is the (1-D) tangential velocity dispersion, yields the following expression
for $\sigma_t(r)$
$$ \sigma_t = \left(\frac{V_{\rm
c}^2}{2}-\sigma_r^2\frac{(\alpha-2)}{2} -\frac{1}{2\pi}\frac{r}{l}
\frac{\sigma_+^2}{(1+[(r-r_0)/l]^2)}\right)^{1/2}~~, \eqno(4)$$
where $\alpha$ is the BHBF star density power-law index defined above.
The line-of-sight velocity dispersion of a set of stars located at a distance $d$
in a field at Galactic
coordinates $(l,b)$ where the velocity ellipsoid has components $\sigma_r$ and
$\sigma_t$ is $$\sigma_{\rm los} =
\sqrt{\gamma^2\sigma_r^2+(1-\gamma^2)\sigma_t^2}\;, \eqno(5)$$ 
where $\gamma$ is a 
simple geometric projection factor $$ \gamma \equiv (d - {r_\odot}{\rm
cos}\,l\,{\rm cos}\,b)/r ~~. \eqno(6)$$ 

\begin{figure}
\psfig{file=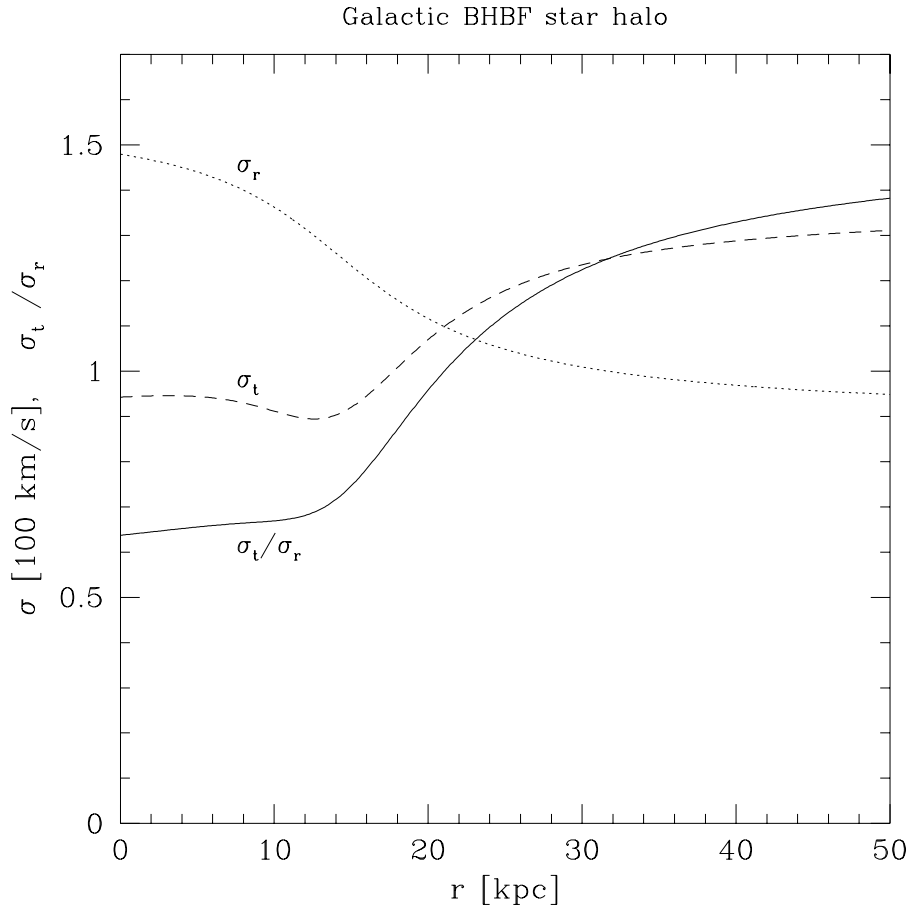,height=12cm,width=12cm}
\caption[]{Best-fit model. The dotted line shows the radial velocity
dispersion $\sigma_r(r)$ and the dashed line shows the tangential velocity
dispersion $\sigma_t(r)$. Also shown is $\sigma_t/\sigma_r$ as a function
of $r$ (solid line).}
\end{figure}

The parameters $(\sigma_0,\sigma_+, r_0, l)$ are determined by maximum
likelihood fitting of expressions
[1] and [4] to the data using eq. [5]. A surprisingly good fit is obtained for this simple
model - see Sommer-Larsen {\it et al.} The resulting $\sigma_r(r)$, $\sigma_t(r)$
and $\eta(r) \equiv \sigma_r(r)/\sigma_t(r)$ are shown in Figure 1. The 
asymptotic value of $\sigma_r(r)$ at large $r$ is found to be
$\sigma_0 = 89 \pm 19$ \kms. 

The main result of the analysis is that $\sigma_r$ decreases from 140--150 \kms at 
$r = r_{\odot}$ to $89\pm 19$ \kms at large $r$ - matched by
a corresponding increase in $\sigma_t$. Hence, the most important kinematic
feature of the model is that the velocity ellipsoid changes from {\it radial}
anisotropy in the solar vicinity ($\eta \simeq$ 0.65) to {\it tangential} anisotropy in 
the outer halo ($\eta \sim$ 1.4).

\section{Outer stellar halo kinematics: clues about the formation of the Milky Way}

The results concerning the dynamics and kinematics of the outer stellar halo are
of considerable interest in relation to theories of the formation of the
Milky Way, in particular, and galaxies in general:

If the Galaxy formed from a
single collapsing over-dense region in the early universe, then one might 
expect the outer halo to be characterized by radially anisotropic 
kinematics (see, e.g., van Albada
1982), whereas the data show that quite the opposite is the case. If, on the
other hand, at least the outer parts of 
the proto-Galaxy were assembled by accretion of small subsystems, then
a large tangential velocity dispersion in the outer parts of the Galaxy is
possible, depending on the nature of the accretion (Norris 1994; Freeman 1996).
So the results indicate that the outer stellar halo formed by some sort of accretion 
and merging processes. The kinematics of stars in the inner halo are, at least locally,
radially
anisotropic, possibly indicating that the inner parts of the halo formed during
a more dissipative and coherent collapse.

Chemical evolution arguments lead to a similar conclusion on the basis of the finding
that there is a significant abundance gradient in the inner halo, but essentially none
in the outer halo. This was discussed in the pioneering work by Searle \& Zinn (1978)
and in much subsequent work --- see, e.g., Norris (1996) and references therein.

\begin{figure}
\psfig{file=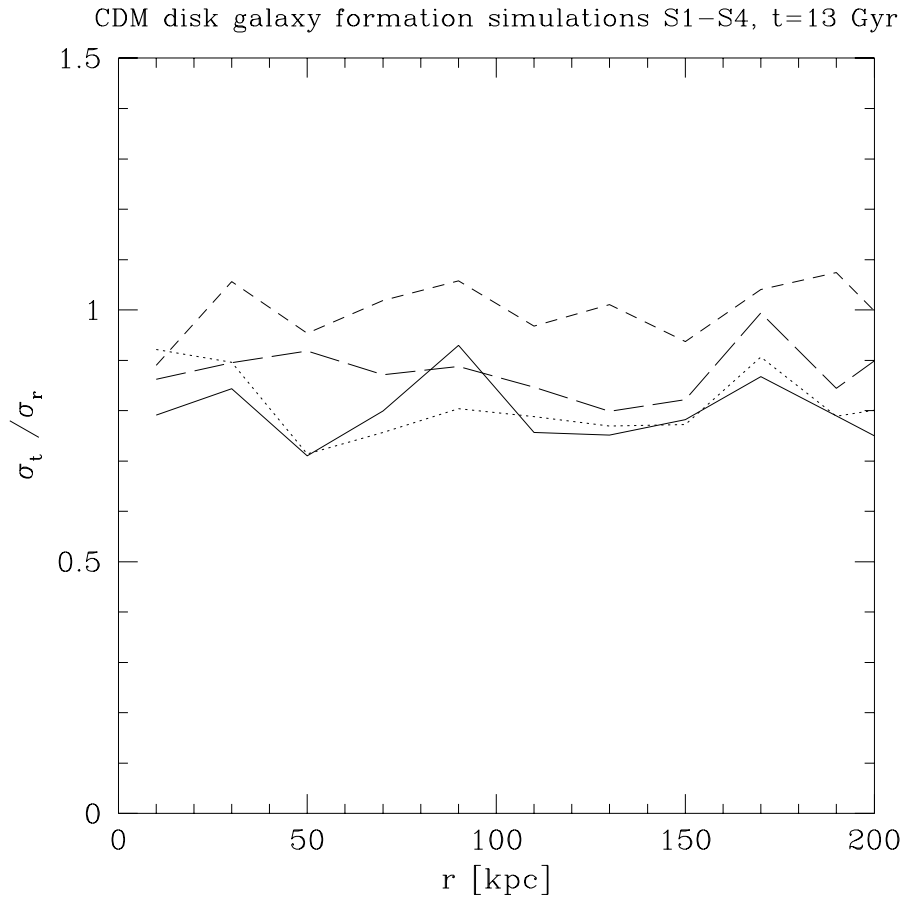,height=12cm,width=12cm}
\caption[]{$\sigma_t/\sigma_r$ as a function of $r$ for the dark matter haloes
of four Milky Way sized model galaxies}
\end{figure}

\begin{figure}
\psfig{file=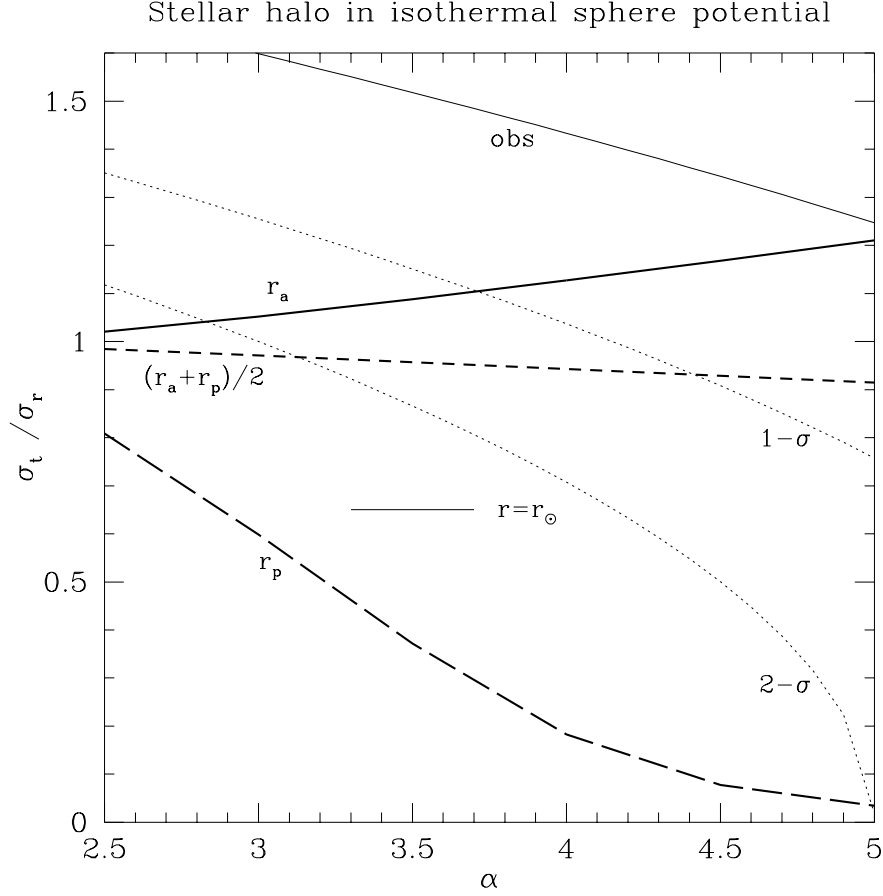,height=12cm,width=12cm}
\caption[]{$\sigma_t/\sigma_r$ as a function of power-law index $\alpha$ 
for models and observations - see text for details}
\end{figure}

In the following I will attempt to quantify the connection to galaxy formation theory,
more specifically the Cold Dark Matter (CDM), hierarchical galaxy formation
scenario: 

Sommer-Larsen, Gelato \& Vedel (1998) carried out cosmological (CDM), 
gravitational/hydrodynamical,
Tree-SPH simulations of the formation and evolution of large (Milky Way sized)
disk galaxies. For the dark matter haloes of four different model galaxies (at the present epoch)
the ratio $\eta(r)$ is shown in Figure 2. As can be seen from the
Figure $\eta \simeq 0.9 \pm 0.1$. This is  
quite different from what is found in the outer stellar halo of the Milky Way and
close to isotropic. Furthermore the rotation curves of the model galaxies are 
approximately flat over the range $r \sim$ 10--100 kpc.
Hence it should be a reasonable approximation for the present purpose to represent
a dark matter halo by an isothermal sphere with phase-space
distribution function
$$
f_{DM}(E) \propto \exp\left(-\frac{\vec{v}^2 + 2 \Phi(r)}{V_c^2}\right) ~~, \eqno(7)
$$
where $\Phi(r)$ is the gravitational potential given by eq. [2].  The dark matter
density falls of like $\rho_{DM} \propto r^{-2}$, whereas the stellar halo
density profile is considerably steeper, $\alpha \simeq 3.4$, so the halo star
formation efficiency $\epsilon_{\star}$ must have had a dependence on some
radial property $r_{orb}$ of the dark matter orbits such that
$$
\epsilon_{\star} \propto r_{orb}^{2-\alpha} ~~. \eqno(8)
$$
One can then ``build'' the stellar halo using dark matter orbits weighted by
$\epsilon_{\star}$. If the halo stars were formed in small, proto-galactic
subsystems before or as they were accreted onto the main dark matter
halo it would be reasonable to take $r_{orb} = r_a$, where $r_a$ is the
apocenter distance of the orbit in the main dark matter halo. If, on the
other hand, the halo star-formation was triggered by disk and halo shocking
of the gas in the subsystems when these were near the inner turning points
of their orbits
one would take $r_{orb} = r_p$, where $r_p$
is the pericenter distance of the orbit. Finally one might also study an
intermediate case like $r_{orb} = (r_a + r_p)/2$. Figure 3 shows the resulting 
values of $\eta$ as a function of the power-law
index $\alpha$ of the halo stars for $r_{orb} = r_a$ (thick solid line),
$r_{orb} = r_p$ (thick long-dashed line) and $r_{orb} = (r_a + r_p)/2$ (thick
short-dashed line) --- $\eta$ does not depend on $r$ for 
the models I consider here since these are scale-free.
Also shown is the asymptotic value (at large $r$) of $\eta$ from the 
observations of the outer
halo (solid line) and 1 $\sigma$ and 2 $\sigma$ deviations (dotted lines).
Finally the local value of $\eta \simeq$ 0.65 is indicated by
a short solid line. As can be seen from the Figure the model predictions do
not agree well with the observations: for $r_{orb} = r_a$ the model 
predictions can be rejected with 82\% confidence for $\alpha$ = 3.4 and the
$r_{orb} = r_p$ case can almost completely be excluded. Hence it appears
unlikely that the formation of the halo stars was
controlled by disk and halo shocking, at least in the framework of the models
presented here.
More observations of BHBF stars in the outer part of the Galactic halo
($r \ga$ 30--40 kpc) are needed in order to reduce the statistical uncertainties, 
but outer stellar halo kinematics clearly has the potential of becoming  
an important constraint on viable galaxy formation scenarios.

\section{Acknowledgements}
I have benefited from discussions with Tim Beers, Per Rex Christensen, Chris Flynn,
Ken Freeman, Sergio Gelato, Steve Majewski, John Norris and Bernard Pagel.

{}

\end{document}